\documentstyle[12pt]{article}
\begin{document}
\title{
\vspace{-11mm} {\normalsize
\begin{tabbing}
\'DESY 92--179 \`hep-th/9212099\\
Dezember 1992
\end{tabbing} \vspace{7mm}}
The Tachyon in a Linear Expanding Universe}
\author{K. Behrndt\thanks{e-mail: behrndt@znsun1.ifh.de}\\
{\normalsize \em DESY-Institut f\"ur Hochenergiephysik, Zeuthen}}
\date{}
\maketitle
\begin{abstract}
We investigate the tachyon coupling in a static Robertson--Walker like
metric background. For a tachyon and dilaton field which are only time
dependent one can rewrite this model as a SU(2) Wess--Zumino--Witten model
and a scalar Feigin--Fuchs theory.
In this case the restriction to a real exponential tachyon field fixes
the level $k$ of the Wess--Zumino--Witten model.
For a spatially dependent tachyon the world radius and the dilaton
are quantized in terms of $k$ and the tachyon by two integers, i.e.
one has a discrete set of fields. The spatial part of the tachyon is
given by Chebyshev polynomials of the second kind. An investigation
of the tachyon mass shows that the tachyon is massless for $k=1$.
\end{abstract}

\renewcommand{\arraystretch}{1.6}
\renewcommand{\thefootnote}{\alph{footnote}}

Strings in cosmological background were discussed a lot in the last
years. Mainly, there are two different approaches: 1) in terms of (gauged)
Wess--Zumino--Witten (WZW) models yielding an exact 2d conformal field
theory \cite{lust}; 2) via the $\sigma$ model (or effective action) approach
in which one gets results in the $\alpha'$ expansion \cite{myer,13}.
In this paper we want to make some general remarks about the description
of strings in a Robertson Walker universe especially for
$\epsilon = +1$ (for $\epsilon = 0$ a general solution
was found by Mueller \cite{Mull}).
The crucial point in this description is the fact
that a Robertson--Walker (RW) space time is conformally flat and therefore it
is always possible to find a coordinate system in which: $G_{\mu\nu}=
\frac{W(r)}{r^2} \eta_{\mu\nu}$, where the scale factor
$W(r)$ describes the original time dependence of the RW metric.
One possibility to handle models like this is given by the world sheet
$\sigma$--model in which one looks for special solutions
for the vanishing of the Weyl anomaly.

After some general remarks about the $\sigma$--model approach we discuss
the description of strings in RW space time.
As one example we investigate a linear expanding universe in four space time
dimensions. As it is known this model corresponds to a
combined SU(2) WZW and Feigin--Fuchs theory \cite{elis,khur}.
Our special interest in this model is to find out what  the
tachyon field looks like which reproduces in the flat limit the known
results from the David--Distler--Kawai (DDK) model \cite{8}
(in general supplemented by spatial background charges and by the
interpretation of the time as Liouville field \cite{3}).
Similar to this model the demand for a real exponential
tachyon yields a restriction: if the tachyon is only time
dependent the level of the WZW theory must be equal to one. Finally
we discuss a spatial dependence and the mass of the tachyon field and
find a discrete set of fields where the spatial part is given
by the Chebyshev polynomials of the second kind.

\vspace{5mm}

The usual $\sigma$--model containing the massless modes and the
tachyon field is given by:
\begin{equation}
  \begin{array}{l}
    Z = \int D X ~ e^{-S} ~~, \\
    S = \frac{1}{4\pi\alpha'} \int_M d^2 z \sqrt{g}\left(g^{ab}
     \partial_a X^{\mu} \partial_b X^{\nu} G_{\mu\nu} +
  i\frac{\epsilon^{ab}}{\sqrt{g}} \partial_a X^{\mu} \partial_b
    X^{\nu}  B_{\mu\nu} + \alpha' R^{(2)} \phi
     + \alpha' T \right)~.
  \end{array}
\end{equation}
Here $G_{\mu\nu}$ corresponds to the metric in the target space
(space time), $B_{\mu\nu}$ is the antisymmetric tensor field, $\phi$
is the dilaton and $T$ the tachyon field. If one wants to
describe non-critical string theory one can e.g.~interpret
the time as Liouville field. In this case the 2d metric
$g_{ab}$ is a reference metric.  The vanishing of the
Weyl anomaly gives the known restrictions for the background
fields: $G, B, \phi$ and $T$ \cite{tseyt2}:
\begin{equation}
 \begin{array}{llr}
 & \bar{\beta}^i \equiv 0 ~~~~\forall i~~ &\mbox{( i numerates all
                                    background fields )} , \\
\mbox{with}:
 \quad& \bar{\beta}^{G}_{\mu\nu} =
                  \beta^{G}_{\mu\nu} + D_{(\mu}M_{\nu)}
                 ~~,~~& M_{\nu} = 2\alpha' \partial_{\nu} \phi + W_{\nu}~,\\
  & \bar{\beta}^{B}_{\mu\nu}= \beta^B_{\mu\nu} + H_{\mu\nu\lambda}
       M^{\lambda} + \partial_{[\mu} K_{\nu]}~,\\
  &  \bar{\beta}^{\phi} = \beta^{\phi} +
                          \frac{1}{2} M^{\mu}\partial_{\mu}\phi~, &\\
  &  \bar{\beta}^T = \beta^T - 2T + \frac{1}{2} M^{\mu}\partial_{\mu}T ~. &
 \end{array}
\end{equation}
Up to the second order in $\alpha'$ one gets for the
$\beta$ functions \footnote{I neglect in my
consideration all `` non-perturbative '' contributions \cite{11}.}:
\begin{equation}
 \begin{array}{rcl}
  \beta^T&=&-\frac{1}{2} \alpha' D^2 T - \alpha'^2\frac{1}{8}
      (H^2)^{\mu\nu} D_{\mu}\partial_{\nu}T ~, \\
  \beta^{G}_{\mu\nu}&=&\alpha' \hat{R}_{(\mu\nu)} + \frac{1}{2} \alpha'^2
      \left(\hat{R}^{\alpha\beta\lambda}_{~~~(\nu}\hat{R}_{\mu)\alpha
      \beta\lambda} -\frac{1}{2} \hat{R}^{\beta\lambda\alpha}_{~~~(\nu}
      \hat{R}_{\mu)\alpha\beta\lambda} +\frac{1}{2}
      \hat{R}_{\lambda(\mu\nu)\beta} (H^2)^{\lambda\beta}\right)~,\\
  \beta^{B}_{\mu\nu}&=&\alpha' \hat{R}_{[\mu\nu]} + \frac{1}{2} \alpha'^2
      \left(\hat{R}^{\alpha\beta\lambda}_{~~~[\nu}\hat{R}_{\mu]\alpha
      \beta\lambda} -\frac{1}{2} \hat{R}^{\beta\lambda\alpha}_{~~~[\nu}
      \hat{R}_{\mu]\alpha\beta\lambda} +\frac{1}{2}
      \hat{R}_{\lambda[\mu\nu]\beta} (H^2)^{\lambda\beta}\right)~,\\
  \beta^{\phi}&=&\frac{1}{6}(D-26) - \frac{1}{2}\alpha' D^2 \phi -
      \frac{1}{8} \alpha'^{2} (H^2)^{\mu\nu}D_{\mu}D_{\nu}
      \phi + \\
    & & +\frac{1}{16}\alpha'^2\left(R_{\mu\nu\alpha\lambda}^2 -
         \frac{11}{2} R H H +
      \frac{5}{24}H^4 + \frac{3}{8}(H^2_{\mu\nu})^2 + \frac{4}{3}
      D H \cdot D H  \right) ~.
 \end{array}
\end{equation}
and $W_{\mu} = -(\alpha'^2 /24) \partial_{\mu}H^2$, $K_{\mu}={\cal O}
(\alpha'^3)$, $H_{\mu\nu\lambda}=\partial_{[\mu}B_{\nu\lambda]}$,
$D H \cdot D H \equiv D_{\mu}H_{\nu\lambda\beta} D^{\mu}
H^{\nu\lambda\beta}$
and $\hat{R}_{\mu\nu\lambda\beta}$ is the generalized curvature tensor
computed in terms of the connection: $\hat{\Gamma}^{\mu}_{~\nu\lambda}
 = \Gamma^{\mu}_{~\nu\lambda} - \frac{1}{2} H^{\mu}_{~\nu\lambda}$.
One should note at this point that some terms depends on the
renormalization scheme (see \cite{tseyt2}).
\vspace{5mm} \newline
Let us give two general remarks about these equations. \vspace{2mm} \newline
1. For $B_{\mu\nu}=0$ and up to ${\cal O}(\alpha'^2)$ the tachyon and
dilaton $\bar{\beta}$ functions are given by:
\begin{equation}
 \begin{array}{l}
  \bar{\beta}^T = -\frac{1}{2} \alpha' D^2 T - 2T + \alpha'
   \partial^{\mu} \phi \partial_{\mu}T ~, \\
  \bar{\beta}^{\phi} = \frac{1}{6}(D-26) - \frac{1}{2}\alpha' D^2 \phi
   + \frac{1}{16} \alpha'^2 R^2_{\mu\nu\alpha\lambda} +
   \alpha' \partial^{\mu} \phi \partial_{\mu}\phi~.
\end{array}
\end{equation}
After the field redefinition:
$T = e^{\phi} \tilde{T}$, $e^{-2 \phi} = f$ and using the vanishing
of the metric $\bar{\beta}$
function the dilaton and tachyon field decouple and
we get the equations:
\begin{equation}
 \begin{array}{l}
  -\frac{1}{2} \alpha' D^2 \tilde{T} - \left( \frac{D-2}{12} - \frac{1}{8}
   \sqrt{\alpha'}R - \frac{1}{32} \alpha' R^2
   + ... \right)\tilde{T} = 0~, \\
  - \frac{1}{2}\alpha' D^2 f + \left( \frac{26-D}{6} + \frac{1}{16}
  \alpha'^2 R^2 +... \right)f = 0 ~.
\end{array}
\end{equation}
We see that both fields fulfill a Klein--Gordon equation of motion where
the mass depends on the dimension of the space time and on
curvature terms. In the flat limit we get just the known
result
that the tachyon corresponds to a massless field for $D=2$ and the
dilaton to a massless field for $D=26$ (if one interprets the time
as Liouville field than $D$ contains the Liouville degree of freedom too).
\vspace{2mm} \newline
2. Beside the decoupling of the tachyon and dilaton fields there
is another decoupling which one should discuss at this point.
The effective action which corresponds to the equations (2)
is given by \cite{tseyt2}
\footnote
{In order to get the tachyon $\bar{\beta}$ function from the effective
action it is necessary to include non-perturbative terms in the
$\bar{\beta}$ functions which we want to neglect in our consideration
(see \cite{11}). The reason is that a (covariant) tachyon coupling in
the effective action would (via the equation of motion) influences
the metric. But tachyon terms in the metric $\beta$ function
arise only via non--perturbative contributions. }:
\begin{equation}
S_{eff} \sim \int d^{D}X
\sqrt{G} e^{-2\phi} \left[ \frac{2}{3}(26-D) + \alpha' (R + 4\alpha'
(\partial \phi)^2 - \frac{1}{12} H^2_{\lambda\mu\nu}) + \alpha'^2
\frac{1}{4} R^2_{\lambda\mu\nu\beta} + ... \right]~.
\end{equation}
In this ``$\sigma$--model parametrization'' the graviton and dilaton
mix in the propagator. After the Weyl transformation and and rescaling
of the dilaton:
\begin{equation}
  \tilde{G}_{\mu\nu} = e^{-\frac{4}{D-2}\phi} G_{\mu\nu} ~~~,~~~
  \tilde{\phi} = \sqrt{\frac{2}{|D-2|}} \phi
\end{equation}
we get the effective action in the ``$S$--matrix parametrization''
containing the ``right'' kinetic terms:
\begin{equation}
S_{eff} \sim \int d^{D} X \sqrt{\tilde{G}} \left[
   \alpha' \tilde{R} - \alpha'
   (\partial \tilde{\phi})^2 +
   \frac{2}{3}(26-D) e^{\frac{4}{\sqrt{2|D-2|}}\tilde{\phi}}
   - \alpha'\frac{1}{12} H^2_{\lambda\mu\nu}
  e^{-\frac{8}{\sqrt{2|D-2|}}\tilde{\phi}} + ... \right]~.
\end{equation}
The equations of motion following from this effective action
are in the lowest order given by \cite{call1}:
\begin{equation}
\begin{array}{c}
\tilde{R}_{\mu\nu} -\frac{1}{2} \tilde{G}_{\mu\nu} \tilde{R} =
  T_{\mu\nu}^{matter}~,\\
 3 \sqrt{2|D-2|} \alpha' D^2 \tilde{\phi} + 4(26-D)
    e^{\frac{4}{\sqrt{2|D-2|}}\tilde{\phi}} + \alpha' H^2_{\lambda\mu\nu}
  e^{-\frac{8}{\sqrt{2|D-2|}}\tilde{\phi}} = 0~,\\
   D_{\lambda}[e^{-\frac{8}{\sqrt{2|D-2|}}\tilde{\phi}}
  H^{\lambda}_{~\mu\nu}] = 0
\end{array}
\end{equation}
where:
\begin{equation}
T_{\mu\nu}^{matter} = \frac{1}{4} \left( H^2_{\mu\nu} - \frac{1}{6}
  \tilde{G}_{\mu\nu} H^2 \right) e^{-\frac{8}{\sqrt{2|D-2|}}\tilde{\phi}}
+  \left( \partial_{\mu}\tilde{\phi} \partial_{\nu}\tilde{\phi}
 - \frac{1}{2} (\partial \tilde{\phi} )^2 \tilde{G}_{\mu\nu} \right)~.
\end{equation}
Apparently one obtains in the limit $\alpha'\rightarrow 0$ just the
general relativity. There is another reason which prefers this
parametrization. Due to the mixing of the
dilaton and graviton the vertex operators following from the
``$\sigma$--model parametrization'' are not well defined (see e.g.
\cite{tseyt4}), e.g.
the dilaton vertex operator has not the right conformal behaviour.
Therefore it is common to consider not $G_{\mu\nu}$ as the
``physical'' metric but $\tilde{G}_{\mu\nu}$.
Later, in our consideration
this reparametrization yields the time evolution of the metric.
\vspace{3mm} \newline
To make us more familiar with this framework let us shortly
rederive known result in flat space time  without Torsion
($G_{\mu\nu}= \eta_{\mu\nu}$, $B_{\mu\nu}=0$).
In this case one has:
\begin{equation}
 \bar{\beta}^{G}_{\mu\nu} = 2\alpha' \partial_{\mu}\partial_{\nu} \phi = 0
\end{equation}
and hence
\begin{equation}
 \phi = \phi_0 - \frac{1}{\sqrt{\alpha'}} q_{\mu} X^{\mu}~,
\end{equation}
which was already discussed as one example for noncritical strings
in arbitrary dimensions \cite{myer}. The tachyon field is defined by:
\begin{equation}
 \bar{\beta}^T = - \frac{1}{2}\alpha'\partial^2 T - 2T - \sqrt{\alpha'}
      q_{\mu}\partial^{\mu}T = 0 ~,
\end{equation}
with the solution:
\begin{equation}
 T \sim e^{-\frac{1}{\sqrt{\alpha'}}p_{\mu}X^{\mu}} \hspace{10mm}
  \mbox{and:} \hspace{10mm}
  -\frac{1}{2}p_{\mu}p^{\mu} - 2 + q_{\mu}p^{\mu} = 0~.
\end{equation}
For the dilaton $\bar{\beta}$ function one gets:
\begin{equation}
 \bar{\beta}^{\phi} = \frac{1}{6}(D-26) + q_{\mu}q^{\mu} = 0 ~.
\end{equation}
Combining (14) and (15) one obtains finally the known result:
\begin{equation}
 \begin{array}{l}
   q_0 = \sqrt{\frac{26-D}{6} - \vec{q}^{\;2}} ~, \\
 p_{0_{\pm}} = q_0 \pm  \sqrt{q_0^{2} - \vec{p}^{\;2} + 2\vec{q}\vec{p} - 4} =
   \frac{1}{\sqrt{6}} \left( \sqrt{26-D - 6\vec{q}^{\;2} } \pm
   \sqrt{ 2-D - 6(\vec{q} - \vec{p})^2 } \right) ~.
 \end{array}
\end{equation}
In the limit $\vec{q}=\vec{p}=0$ and for $D=d+1$ ( spatial degrees
of freedom plus Liouville field) we get just the known results from
David, Distler and Kawai \cite{8} (it is common to replace: $p_{\mu}
\rightarrow i p_{\mu}$ and $q_{\mu} \rightarrow i q_{\mu}$).

\vspace{5mm}

Now we want to investigate a (euclidean) Robertson-Walker
space time. This metric describing a (spatial) homogeneous
and isotropic universe is given by:
\begin{equation}
  (ds)^2 = (dt)^{2} + \frac{K^2 (t)}{(1 +
\frac{1}{4}\epsilon \bar{r}^2)^2}
       \left[ (dX^1)^2 + (dX^2)^2 + ... + (dX^{D-1})^2 \right] ~,
\end{equation}
where $\bar{r}^2 = (X^1)^2 + (X^2)^2 + ... + (X^{D-1})^2$ ,
$K(t)$ is the so-called
world radius and the parameter $\epsilon$ determinates the spatial geometry:
flat ($\epsilon = 0$), spherical ($\epsilon = +1$) or hyperbolical
($\epsilon = -1$).
Furthermore we restrict ourselves to the compact case $\epsilon = +1$.
Since all RW metrics are conformally flat we can find coordinates
in which the metric looks like\footnote{Of course,
one can not find a {\em global} transformation which transforms
the compact space  to a non-compact space but for every
finite part it is possible.}:
\begin{equation}
(ds)^2=\frac{\tilde{K}^2 (r)}{r^2}\left( (dx^0)^2 + (dx^1)^2 + ...
      + (dx^{D-1})^2 \right)~.
\end{equation}
where the new radius $r^2 = (x^0)^2 + |\vec{r}|^2$ and the new
time $x^0$ depends on $\bar{r}$ and $t^0$ via:
\begin{equation}
\begin{array}{lll}
  |\vec{r}| = e^{\eta} r_1 & \quad \mbox{and} \quad &
   r_1 = \frac{\bar{r}}{1+\frac{1}{4} \bar{r}^2} \\
  x^0 = e^{\eta} \sqrt{1- r_{1}^{2}} & &
  \eta = \int \frac{d t}{K}~.
\end{array}
\end{equation}
Hence the new radius $r$ corresponds just to the former time:
$\log r =\eta$ and one obtains the original world radius $K$ from
$\tilde{K}$ after the
replacement of $r$ by $t$ where $r(t)$ is given by:
\begin{equation}
\frac{\dot{r}}{r}=\frac{1}{\tilde{K}(r)}~.
\end{equation}
Thus the whole time dependence of the original
metric is now controlled (via the radius dependence) by the
Weyl factor $\tilde{K}$.
Let us give some simple examples:
\begin{equation}
\begin{array}{lll}
1.\qquad \tilde{K}=\sqrt{Q}=const.&\rightarrow&static~Einstein~universe\\
2.\qquad \tilde{K}=Qr~(\mbox{local~flat~space})
   &\rightarrow&        K=t \quad: Milne~universe\\
3.\qquad \tilde{K}=Qr^{n} &\rightarrow&K=n t \quad : general~linear~
       exp.~universe\\
4.\qquad \tilde{K} \sim \frac{r}{1+\frac{1}{4}r^2}&\rightarrow&
  K \sim \sin t\quad : de~Sitter~universe
\end {array}
\end{equation}
In the following we want to discuss the first case for which an exact
solution in four dimensions exists \cite{elis,khur}. The main point to get
the exact result is to rewrite the model (1) in a WZW model
and 1-dim. Feigin--Fuchs model. This can be done under the assumption
that $D=4$, the tachyon and dilaton field depend on $r$
only, i.e.~in the old coordinates (17) they are only time dependent.
In a second step we turn to the investigation of a more general tachyon
field in this background.

If:
\begin{equation}
\begin{array}{l}
G_{\mu\nu}=\frac{Q}{r^2}\delta_{\mu\nu} \qquad \qquad \mu,\nu = 0,1,2,3,\\
H_{\mu\nu\lambda}=\pm\epsilon_{\mu\nu\lambda}^{~~~~\sigma}
  \partial_{\sigma}\log\frac{\sqrt{Q}}{r}\\
\phi = \phi(r) \quad \mbox{and}\quad T=T(r)
\end{array}
\end{equation}
we can write (1) as:
\begin{equation}
\begin{array}{l}
S = S_{WZW}(SU(2)) + \frac{1}{4\pi\alpha'}\int\left( (\partial u)^2 Q
    + \alpha' R^{(2)} \phi(u) + \alpha' T(u) \right)\\
\mbox{and}\\
S_{WZW}(SU(2)) = \frac{Q}{4\pi\alpha'}\int d^2 z tr\left( \partial_a g^{-1}
   \partial_a g \right) + \frac{Q}{4\pi\alpha'}\int d^3 z tr \epsilon^{abc}
   \left( (g^{-1}\partial_a g) (g^{-1} \partial_b g) (g^{-1} \partial_c
   g) \right)
\end{array}
\end{equation}
where $g = \frac{1}{r}(x^0 {\bf 1} +
x^{i}${\bf $\sigma^i$}), $u=-\log r$ ($\sigma^i$:
Pauli matrices).
If one introduces polar coordinates in $g$ one can see
that the special background fields (22) just decouples the
dependence on the radius $r$ from the dependence on the angles.
The dependence on the angles is now controlled by the WZW action
and the dependence on the radius by the Feigin--Fuchs part.
Background fields like (22) are already discussed in the past, e.g.~in
the context of semi--wormhole or solitons in string theory \cite{strom}
and the torsion as a special axion field.
The WZW theory is well defined and conformal invariant if \cite{witt}:
$\frac{Q}{\alpha'} = k = 1, 2, 3, ...$ is the WZW level and the corresponding
central charge for the $SU(2)$ Kac--Moody algebra is given
by: $c_{wzw} = \frac{3k}{k+2}$.
In order to get a conformal theory the dilaton and the tachyon
field have to satisfy the (corresponding $\bar{\beta}$--function)
equations:
\begin{equation}
\begin{array}{c}
\phi''(u)=0\\
1+c_{wzw}-26 + 6 \alpha' (\phi ')^2=0\\
-\frac{1}{2} \alpha' T'' - 2T + \alpha' \phi' T' = 0
\end{array}
\end{equation}
($26$ is the contribution from the ghosts to the central charge).
If the tachyon and dilaton satisfy these equations we have an exact
conformal theory, i.e. in all (finit) order in $\alpha'$. From
the first equation (corresponds to the metric $\bar{\beta}$ function)
follows that $\phi$ is at most linear in $u$, the second
equation ensures the vanishing of the total central charge
and the last equation defines the tachyon field. As
solution one finds:
\begin{equation}
\begin{array}{lll}
&\phi(u)=\phi_0 + q u \qquad \mbox{with}:&
 q^2 = \frac{k}{6}(25 - \frac{3k}{k+2})\\
k>1:&
T_{\pm}(u)\sim e^{\alpha_{\pm} u} &
  \alpha_{\pm} = \sqrt{k/6}\left(\sqrt{25-\frac{3k}{k+2}}\pm
                    \sqrt{1-\frac{3k}{k+2}}\right)\\
k=1:& T_1(u) \sim e^{2u} \quad ; \quad T_2(u) \sim u e^{2u}& .
\end{array}
\end{equation}
The importance of the second tachyon operator $T_2$ was already
discussed in \cite{polch}. Especially, $T_2$ plays a crucial rule
for the explanation of the critical behaviour in comparison to
the results coming from random matrix model. Since for $k>1$ $\alpha_{\pm}$
becomes complex $T_{\pm}$ correspond to oscillating (real) fields:
 $T(u) \sim e^{q u} (\frac{3k}{k+2}-1)^{-\frac{1}{2}}
\sin (\sqrt{\frac{k}{6}(\frac{3k}{k+2}-1)}u)$ (and a corresponding
solution with $\cos(\sqrt{\frac{k}{6}(\frac{3k}{k+2}-1)} u)$.
If one wants to have a real exponential tachyon (and not an
oscillating one) motivated by
obtaining the known results in non-critical string theory
in the flat limit \cite{8} one gets the restriction: $k=1$.
In addition, $k=1$ is prefered because this corresponds to a massless
tachyon field (see below).
Transforming the background fields (25) and (22) back in the coordinate
system (17) and denoting the integration constants as a suitable
$t_0$ we get:
\begin{equation}
\begin{array}{l}
(ds)^2 = dt^2 + \frac{\alpha'}{(1+\frac{1}{4} \bar{r}^2)^2}
         [(dX^1)^2+(dX^2)^2+(dX^3)^2] \quad , \quad
H_{\mu\nu\lambda}= \pm \epsilon_{\mu\nu\lambda 0} ~,
\\
\phi(t) =-\frac{2}{\sqrt{\alpha'}} (t-t_0) \quad , \quad
T_1 = e^{-\frac{2}{\sqrt{\alpha'}}(t-t_0)} \quad , \quad
T_2 \sim (t-t_0) e^{-\frac{2}{\sqrt{\alpha'}}(t-t_0)} ~.
\end{array}
\end{equation}
We should recall here that this is an exact conformal field theory.
The reason to get this theory was the special form of the
metric and torsion (22) which decouple the degrees of freedom:
angle part $\sim$ WZW theory and radius part $\sim$ Feigin--Fuchs
theory (both are conformal exact models).
After the redefinition (7) and a common time
reparametrization we get for the ``physical'' metric:
\begin{equation}
(\tilde{ds})^2 = (dt)^2 + \frac{2(t-t_0)^2}{(1+\frac{1}{4} \bar{r}^2)^2}
      [(dX^1)^2 + (dX^2)^2 + (dX^3)^2] ~.
\end{equation}
Thus we have as result a linear expanding
universe corresponding to a linear dilaton field and an exponential
tachyon with the classical value $\alpha=2$.
If one does not include the tachyon field in the action (1) or (23)
all values of $k$ are allowed and one has
to replace the factor 2 in front of $t-t_0$ by $q$ in the
dilaton and metric. Then we would have in correspondence
to \cite{elis} the assertion that the world radius and the dilaton
field is quantized by $k = 1, 2, 3,...$ .
The incorporation of a (real exponential) tachyon depending only from
the time restricted us to $k=1$. Unfortunately $k$ is
the perturbation parameter in this theory and if $k=1$ the
perturbation theory breaks down. Up to this point this does not matter
because it is an exact solution of (2). The question is what happens
if one has e.g. a more general tachyon field for which the
$\beta$ function is only perturbatively solvable and known.
To investigate this question we want now to discuss a
tachyon field depending on a angle coordinate too.
In this case up to ${\cal O}(\alpha'^2)$
we have to solve the equation:
\begin{equation}
\bar{\beta}^T = -\frac{1}{2} \alpha' D^2 T + \frac{1}{8} \alpha'^2
    (H^2)^{\mu\nu} D_{\mu}\partial_{\nu} T - 2T + \alpha'
    \partial_{\mu} \phi \partial^{\mu} T = 0
\end{equation}
where $\phi$ and $H$ are given by (22), (25).  The dependence on the
angle we separate by a ansatz which gives in the flat limit
plane waves (see below):
\begin{equation}
T = (r^2)^{-a} V(\frac{p\cdot x}{r}) ~,
\end{equation}
where the coefficient $a$ corresponds to $\alpha$ in (25).
After using of: $x^{\mu} \partial_{\mu}V
= 0$ and $\frac{p \cdot x}{r} = z$ one finds immediately:
\begin{equation}
( |p|^2 - z^2 ) V''(z) - 3z V'(z) + \frac{4}{1 - \frac{1}{2k}}
(a^2 - qa + k) V = 0 ~.
\end{equation}
In order to interpret this equation it is useful to look on
the flat limit.
This can be done by the transformations (see (17),$K^2 = Q$, for dimensional
reasons one has to replace the denominator by $(\alpha'+\frac{1}{4}\epsilon
\bar{r}^2)^2$):
\begin{equation}
\vec{X}~\rightarrow~\lambda \vec{X} \qquad , \qquad Q \rightarrow
\frac{\alpha'}{\lambda^2}  \qquad \mbox{and} \qquad \lambda \rightarrow 0
\end{equation}
which corresponds to $k=\frac{Q}{\alpha'}\rightarrow
\frac{1}{\lambda^2} $, i.e. the flat limit corresponds to
$k\rightarrow \infty$. If one looks firstly on the case $V=const$
corresponding to the solution (25) one finds immediately that
$a={\cal O}(\frac{1}{\lambda})$, $q ={\cal O}(\frac{1}{\lambda})$ and
in terms of $z=q^0+\lambda\frac{\vec{p}\vec{X}}{\sqrt{\alpha'}}+{\cal O}
(\lambda^2)$ we get for (30) in the leading order
(${\cal O}(\frac{1}{\lambda^2})$):
\begin{equation}
\alpha' |\vec{p}|^2 V''(\vec{p} \vec{X}) + 4\left(a^2 - \sqrt{\frac{26-4}
 {6 \alpha'}}a  + 1 \right)V=0
\end{equation}
with the solution: $V \sim e^{-\frac{\vec{p}\vec{X}}{\sqrt{\alpha'}}}$,
$a=\frac{1}{\sqrt{24 }}
\left( \sqrt{26-4}\pm \sqrt{2-4 - 6 |\vec{p}|^2}\right)$. The same
procedure for $T$ and $\phi$ yields: $T=e^{-\frac{2at +
\vec{p}\vec{X}}{\sqrt{\alpha'}}}$ ,
$\phi =-\frac{1}{\sqrt{\alpha'}} q t$. Therefore we get just the known
result (16) if:
\begin{equation}
t \rightarrow X^0 \qquad , \qquad q \rightarrow q_0 \qquad , \qquad
2 a= p_0 \qquad , \qquad
\vec{q}=0 \qquad \mbox{and} \qquad D=4 ~,
\end{equation}
i.e.~$p_{\mu}$ corresponds just to the momentum in the flat limit.
But it is also possible to get an explicit solution for (30).
After the transformations:
\begin{equation}
 z=|p|\cos\theta \qquad (\theta\mbox{ is the angle
 between $p_{\mu}$ and $x^{\mu}$}) \quad \mbox{and} \quad
V(\theta)=\frac{W(\theta)}{\sin \theta}
\end{equation}
one gets for (30):
$W''(\theta) + \rho^2 W(\theta) =0$ and
thus the non-singular solution ($V(\theta =0)=1$) for the tachyon (29)
is given by:
\begin{equation}
T(r,\theta)= r^{p_0} \frac{\sin[\rho \theta]}{\rho \sin\theta} \qquad
, \qquad \rho^2 = \frac{1}{1-\frac{1}{2k}}(p_{0}^2- 2 q p_0 + 4 k -
\frac{1}{2k} +1)
\end{equation}
($q$ given by (25), $p_0$ is the time component of the tachyon
momentum). If we now restrict ourselves to a solution which is
periodic: $V(\theta)=V(\theta + 2\pi)$ we get a quantization
for the momentum $p_0$ \footnote{A similar quantization is obtained
by a suitable normalization.}:
\begin{equation}
\rho = n+1 \quad \leftrightarrow \quad
p_0 = \sqrt{\frac{k}{6}\left(25-\frac{3k}{k+2}\right)} \pm
\sqrt{(n^2 +2n)\left(1-\frac{1}{2k}\right)-\frac{k}{6}
\left(\frac{3k}{k+2}-1\right)}
\end{equation}
where $n$ is an integer. In order to have a real
momentum $p_0$ it is necessary that:
\begin{equation}
|n+1| \geq \frac{1}{\sqrt{1-\frac{1}{2k}}}\sqrt{1-\frac{1}{2k} +
 \frac{k}{6} \left(\frac{3k}{k+2}-1\right)} \quad .
\end{equation}
For $n=0$ we have the restriction $k=1$ and we get the same result
as in (25) or (26).
The final result for a spatially dependent tachyon in the
coordinate system (17) is therefore up to the second order in $\alpha'$
given by:
\begin{equation}
\begin{array}{l}
(ds)^2 = dt^2 + \frac{k \alpha'}{(1+\frac{1}{4} \bar{r})^2}
         [(dX^1)^2+(dX^2)^2+(dX^3)^2] \qquad , \qquad
 H_{\mu\nu\lambda}= \pm \epsilon_{\mu\nu\lambda 0} ~,\\
\phi(t) =\sqrt{\frac{k}{6 \alpha'}(25-\frac{3 k}{k+2})} (t-t_0) \qquad ,
\qquad T(t,\theta)= e^{\frac{p_0 (t-t_0)}{\sqrt{\alpha'}}} \,
\frac{\sin[(n+1)\theta]}{(n+1) \sin \theta}~,\\
\mbox{with :}\qquad
|p| \cos\theta = p_0 \frac{\alpha'-\frac{1}{4}\bar{r}^2}{\alpha'+\frac{1}{4}
\bar{r}^2} + \vec{p}\vec{X}\frac{\sqrt{\alpha'}}{\alpha'+
\frac{1}{4}\bar{r}^2} \quad,
\quad k = 1, 2, 3, ... ~,
\end{array}
\end{equation}
where: $p_0$ is given by (36) and the integer $n$ has to fulfill
(37) (the contribution of the second order is given by term $\frac{1}{2k}$).
We have again a linear dilaton and an exponential time
dependence of the tachyon. The spatially (or $\theta$) dependent part
of the tachyon is given by the Chebyshev polynomials of the second
kind which were already discussed as a representation of SU(2)
current algebra \cite{gepn}.
 The main difference to (26) is that we have now in all fields a
quantization depending on two integers. The world radius and the dilaton
is quantized by $k$ and the tachyon is also quantized by $n$.
A cosmological interpretation of a quantized world
radius is given in the first ref.~of \cite{elis}.
In addition one should note here that the solution (38)
has for even $n$ the duality symmetry: $\bar{r}\rightarrow
-\frac{4\alpha'}{\bar{r}}$ which is equivalent to
$\theta\rightarrow\pi-\theta$.

Finally we want to investigate the mass spectrum of the tachyon
in (26) and (38). As we have seen in (5) the tachyon
and the dilaton corresponds in the lowest order in $\alpha'$
to scalar fields fulfilling a Klein--Gordan equation.
After the field redefinition $T=e^{\phi}\tilde{T}$
($\phi$ and $T$ are given by (38)) we get for the tachyon $\bar{\beta}$
function:
\begin{equation}
\begin{array}{c}
-\frac{1}{2} \alpha' D^2 \tilde{T} - \frac{1}{12}
\left(\frac{3k}{k+2}-1\right) \tilde{T} = 0\\
\mbox{with the solution:}\qquad \tilde{T}(t,\theta) =
e^{\frac{1}{\sqrt{\alpha'}}\tilde{p}_0 (t-t_0)} \frac{\sin[(n+1)\theta]}
 {(n+1)\sin \theta}
\end{array}
\end{equation}
where $\tilde{p}_0 = \pm \sqrt{(n^2 +2n)\left(1-\frac{1}{2k}\right)-\frac{k}{6}
\left(\frac{3k}{k+2}-1\right)}$. Hence, the mass is quantized by $k$
and the tachyon is massless if $k=1$.
Apart from the argumentation
that $k$ has to be one in order to
obtain in the flat limit the known results from the DDK model
the requirement to get a massless tachyon state is an
alternative one (note: the results (25) or (26) for an only time
dependent tachyon correspond to $n=0$ in (38)).
But one has to take into account that this
conclusion is only valid in the low energy limit (neglecting of
higher order in $\alpha'$) because the tachyon $\beta$ function
gets additional terms in the higher orders of $\alpha'$.

\vspace{10mm}

\noindent
{\large \bf Acknowledgments} \vspace{3mm} \newline \noindent
We would like to thank H.~Dorn, D.~L\"ust and M.~Schmidt for
useful discussions.

\renewcommand{\arraystretch}{1}

\end{document}